\documentclass{nature}

\usepackage{graphicx}
\usepackage{setspace}
\usepackage{amsmath}
\usepackage{times}
\usepackage{amssymb}
\usepackage{times}
\usepackage{sectsty}
\usepackage[toc,page]{appendix}

\usepackage{threeparttable}
\usepackage{multirow}
\usepackage{booktabs}

\usepackage{lineno}
\usepackage{color}
\usepackage{xcolor}   
\usepackage{ulem}     

\newcommand{\CCS}{CsCr$_3$Sb$_5$}
\newcommand{\CVS}{CsV$_3$Sb$_5$}

\renewcommand{\footnotesize}{\fontsize{9pt}{9pt}\selectfont}

\def\CCS{CsCr$_3$Sb$_5$}
\def\CVS{CsV$_3$Sb$_5$}

\title{Superconductivity emerging from density-wave-like order in a correlated kagome metal}
\author{Yi Liu$^{1,2\ast}$, Zi-Yi Liu$^{3,4\ast}$, Jin-Ke Bao$^{5,6\ast}$, Peng-Tao Yang$^{3,4}$, Liang-Wen Ji$^{1}$, Si-Qi Wu$^{1}$, Qin-Xin Shen$^{3,4}$, Jun Luo$^{3}$, Jie Yang$^{3}$, Ji-Yong Liu$^{7}$, Chen-Chao Xu$^{5}$, Wu-Zhang Yang$^{8}$, Wan-Li Chai$^{1}$, Jia-Yi Lu$^{1}$, Chang-Chao Liu$^{1}$, Bo-Sen Wang$^{3,4}$, Hao Jiang$^{9}$, Qian Tao$^{1}$, Zhi Ren$^{8}$, Xiao-Feng Xu$^{2}$, Chao Cao$^{1,10}$, Zhu-An Xu$^{1,11,12}$, Rui Zhou$^{3\dag}$, Jin-Guang Cheng$^{3,4\dag}$, Guang-Han Cao$^{1,11,12\dag}$}

\begin{document}

\maketitle

\begin{affiliations}
\item{\normalsize{School of Physics, Zhejiang University, Hangzhou 310058, China}}\
\item{\normalsize{Department of Applied Physics, Key Laboratory of Quantum Precision Measurement of Zhejiang Province, Zhejiang University of Technology, Hangzhou 310023, China}}\
\item{\normalsize{Beijing National Laboratory for Condensed Matter Physics and Institute of Physics, Chinese Academy of Sciences, Beijing 100190, China}}\
\item{\normalsize{School of Physical Sciences, University of Chinese Academy of Sciences, Beijing 100190, China}}\
\item{\normalsize{School of Physics and Hangzhou Key Laboratory of Quantum Matters, Hangzhou Normal University, Hangzhou 311121, China}}\
\item{\normalsize{Department of Physics, Materials Genome Institute and International Center for Quantum and Molecular Structures, Shanghai University, Shanghai 200444, China}}\
\item{\normalsize{Department of Chemistry, Zhejiang University, Hangzhou 310058, China}}\
\item{\normalsize{School of Science, Westlake Institute for Advanced Study, Westlake University, Hangzhou 310064, China}}\
    \item{\normalsize{School of Physics and Optoelectronics, Xiangtan University, Xiangtan 411105, China}}\
\item{\normalsize{Center for Correlated Matter, Zhejiang University, Hangzhou 310058, China}}\
\item{\normalsize{Interdisciplinary Center for Quantum Information, and State Key Laboratory of Silicon and Advanced Semiconductor Materials, Zhejiang University, Hangzhou 310058, China}}\
\item{\normalsize{Collaborative Innovation Centre of Advanced Microstructures, Nanjing University, Nanjing, 210093, China}}\\
{\normalsize{$^\ast$These authors contribute equally to this work.}\\
\normalsize{$\dag$E-mail: ghcao@zju.edu.cn; jgcheng@iphy.ac.cn; rzhou@iphy.ac.cn}}
\end{affiliations}

\begin{abstract}
Unconventional superconductivity (USC)
in a highly correlated kagome system has been theoretically proposed for years
\cite{2009PRB.WXG,2012PRB.LJX,2013PRB.WQH,2013PRL.Kiesel,2017RMP.ZhouY}, yet the experimental realization is hard to achieve\cite{2016RMP.Norman,2016PRX.McQeen}. The recently discovered vanadium-based kagome materials\cite{2019PRM.135}, which
exhibit both superconductivity\cite{2020PRL.Cs135SC,2021PRM.K135,2021CPL.Rb135} and charge density wave (CDW) orders\cite{2021NM.K135.CO,2021PRX.Ortiz,2021PRX.CDW.CXH},
are nonmagnetic\cite{2019PRM.135,2020PRL.Cs135SC} and weakly correlated\cite{2021PRB.weak-corr,2022NC.ARPES}, thus unlikely host USC as theories proposed.
Here we report the discovery of a chromium-based kagome metal, CsCr$_3$Sb$_5$, which is contrastingly characterised by strong electron correlations, frustrated magnetism, and characteristic flat bands close to the Fermi level. Under ambient pressure, it undergoes a concurrent structural and magnetic phase transition at 55 K, accompanying with a stripe-like $4a_0$ structural modulation. At high pressure, the phase transition evolves into two transitions, probably associated with CDW and antiferromagnetic spin-density-wave orderings, respectively. These density-wave (DW)-like orders are gradually suppressed with pressure and, remarkably, a superconducting dome emerges at 3.65-8.0 GPa. The maximum of the superconducting transition temperature, $T_\mathrm{c}^{\mathrm{max}}=$ 6.4 K, appears when the DW-like orders are completely suppressed at 4.2 GPa, and the normal state exhibits a non-Fermi-liquid behaviour, reminiscent of USC  and quantum criticality in iron-based superconductors\cite{2009PRB.Ba122.Canfield,2014review.Matsuda}. Our work offers an unprecedented platform for investigating possible USC in a correlated kagome system.
\end{abstract}

\section{Introduction}\label{sec1}
Materials with two-dimensional kagome lattice are featured with geometric frustration and characteristic electronic structures, from which various intriguing quantum states may emerge\cite{2022nature.review}. Experimental realization of USC (which commonly refers to as non-electron-phonon mediated superconductivity\cite{2013RMP.Scalopino,2017AP.Stewart}) in a kagome lattice is highly valuable\cite{2009PRB.WXG,2012PRB.LJX,2013PRB.WQH,2013PRL.Kiesel,2017RMP.ZhouY,2016RMP.Norman,2016PRX.McQeen}. The vanadium-based kagome materials $A$V$_3$Sb$_5$ ($A$ = K, Rb, Cs) recently discovered\cite{2019PRM.135} present many exotic phenomena including superconductivity\cite{2020PRL.Cs135SC,2021PRM.K135,2021CPL.Rb135,2023nature.Cs135}, unusual charge order\cite{2021NM.K135.CO,2021PRX.Ortiz,2021PRX.CDW.CXH,2022NP.CO.135,2022NP.Cs135,2021nature.Cs135,2022nature.TRSB}, anomalous Hall effect\cite{2020SA.AHE}, pair density wave\cite{2021nature.PDW.GaoHJ}, and electronic nematicity\cite{2022nature.nematicity.CXH}. Nevertheless, this class of materials are most likely phonon-mediated conventional superconductors\cite{2022NSR.review}, in line with the nonmagnetic nature with relatively weak electron correlations\cite{2019PRM.135,2020PRL.Cs135SC,2021PRB.weak-corr,2022NC.ARPES}. Correlated kagome materials,
on the other hand, generally bear robust magnetism hampering the appearance of USC. Below we report a new chromium-based kagome material, CsCr$_3$Sb$_5$, which uniquely hosts both strong electron correlations and fragile DW-like orders. We found that the DW-like orders are gradually suppressed with pressure, and possible USC emerges nearby a quantum critical point (QCP) at $P_\mathrm{c}\approx$ 4 GPa.

\section{Physical properties of CsCr$_3$Sb$_5$}
\begin{figure*}
\includegraphics[width=16cm]{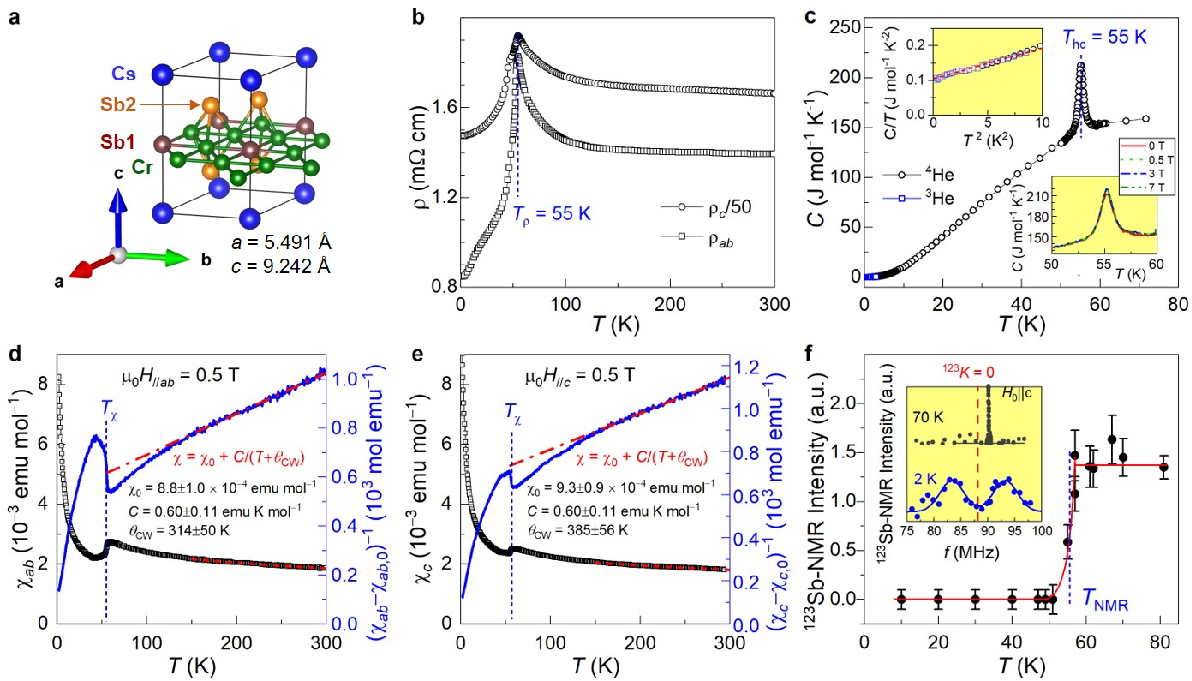}
\caption{\textbf{Crystal structure and physical properties of CsCr$_3$Sb$_5$}. \textbf{a}, Unit cell of the high-temperature hexagonal phase of \CCS. \textbf{b}, In-plane and out-of-plane resistivity as functions of temperature $T$. \textbf{c}, Temperature dependence of specific heat ($C$). The upper inset plots $C/T$ versus $T^2$ at low temperatures, and the bottom inset highlights robustness of the $C(T)$ anomaly against magnetic fields along the $c$ axis.
\textbf{d,e}, Temperature dependence of magnetic susceptibility $\chi$ with magnetic field parallel (\textbf{d}) and perpendicular (\textbf{e}) to the $ab$ plane, respectively. The right axis plots $(\chi-\chi_0)^{-1}$ versus $T$. The red dashed lines denote the Curie-Weiss fit, and the fitted parameters are displayed.
\textbf{f}, Temperature dependence of $^{123}$Sb-NMR signal (corrected for a temperature factor 1/$T$). The inset is the $^{123}$Sb-NMR spectra at 70 and 2 K under an external field of $\mu_0H$ = 16 T along the $c$ axis. The solid lines are the Gaussian fit. The red dashed line marks average of the Knight shift at 2 K. In \textbf{b-f}, anomalies at 55 K associated with the phase transition are marked with blue dashed lines.}
\label{fig:pp}
\end{figure*}

\begin{figure*}
\centering
\includegraphics[width=14cm]{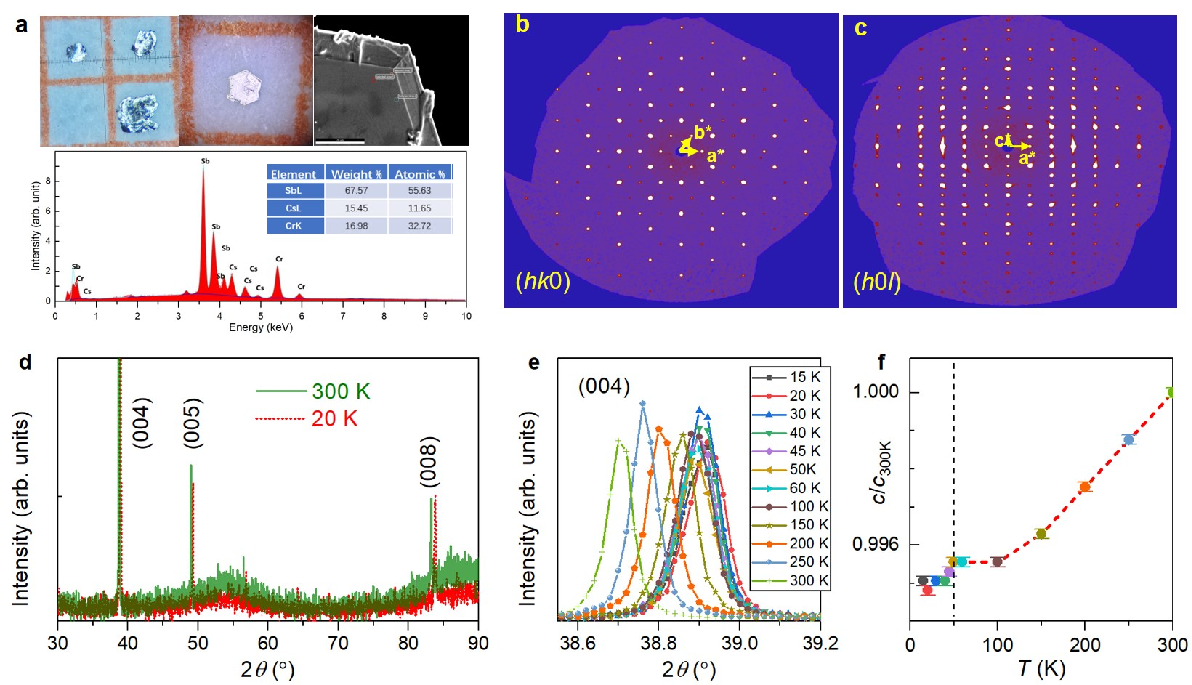}
\caption{\textbf{Extended Data Fig. 1 Characterizations of \CCS~ crystals}. \textbf{a}, Optical photographs (left and middle top), an SEM image (right top), and the typical EDX spectrum. \textbf{b,c}, Reconstructed single-crystal XRD patterns of ($hk0$) and ($h0l$) reflection planes, respectively, at 298 K.
\textbf{d}, XRD $\theta-2\theta$ scan at 300 and 20 K. \textbf{e}, The $(00l)$ reflections at different temperatures. \textbf{f}, Relative lattice parameter $c/c_{\mathrm{300K}}$ as a function of temperature, showing a phase transition at $T=50\pm5$ K.}
\label{fig:Ex_XRD}
\end{figure*}

Single crystals of CsCr$_3$Sb$_5$ were grown via a self-flux method\cite{2020PRL.Cs135SC,2022PRM.LY}. The as-grown crystals were characterized by single-crystal X-ray diffractions (XRD) and energy dispersive X-ray (EDX) spectroscopy (Extended Data Fig.~\ref{fig:Ex_XRD}a). The crystals are typically thin flakes with silver metallic luster and with hexagonal morphology and, furthermore, they are stable in air. The chemical composition is of stoichiometric CsCr$_3$Sb$_5$ within the measurement errors.
The single-crystal XRD at room temperature (Extended Data Figs.~\ref{fig:Ex_XRD}b,c) indicate that CsCr$_3$Sb$_5$ crystallizes in a hexagonal lattice with the space group of $P6/mmm$ (Fig.~\ref{fig:pp}a and Extended Data Table 1). Cr atoms form a two-dimensional (2D) kagome net with Sb1 atoms located at the center of the hexagons. This 2D Cr$_3$Sb plane is sandwiched by the honeycomb-like layers of Sb2, and the resultant sandwiched layers of [Cr$_3$Sb$_5$]$^{-}$ are separated by layers of Cs$^+$ ions. Therefore, CsCr$_3$Sb$_5$ is isostructural to $A$V$_3$Sb$_5$ (ref.\cite{2019PRM.135}).

Figures~\ref{fig:pp}b-f summarise the physical properties of \CCS~at ambient pressure. The in-plane resistivity $\rho_{ab}(T)$ is nearly temperature-independent above $\sim$150 K with an absolute value of resistivity of 1.4 m$\Omega$ cm (Fig.~\ref{fig:pp}b). The resistivity anisotropy $\rho_{c}/\rho_{ab}$ is as high as $\sim$60, indicating a quasi-2D transport property. In the 2D-conduction scenario, the parameter $k_\mathrm{F}l$, where $k_\mathrm{F}$ and $l$ are the Fermi wavevector and mean free path, respectively, is estimated to be close to unity\cite{2010AP.Johnston}, suggestive of a correlated bad metal\cite{1995PRL.badmetal}. The low-$T$ specific-heat data (top inset of Fig.~\ref{fig:pp}c) give information about the strength of electron correlations. The data fitting with the formula $C/T=\gamma+\beta T^2$ yields a large electronic specific-heat coefficient, $\gamma_{\mathrm{exp}}=105(1)$ mJ K$^{-1}$ mol$^{-1}$. This $\gamma$ value is four times larger than that of the V-based \CVS~(ref.\cite{2021SCP.YHQ}), and it is even larger than that of correlated quasi-one-dimensional superconductor K$_2$Cr$_3$As$_3$ (ref.\cite{2015prx.bao}).

\begin{figure*}
\centering
\includegraphics[width=12cm]{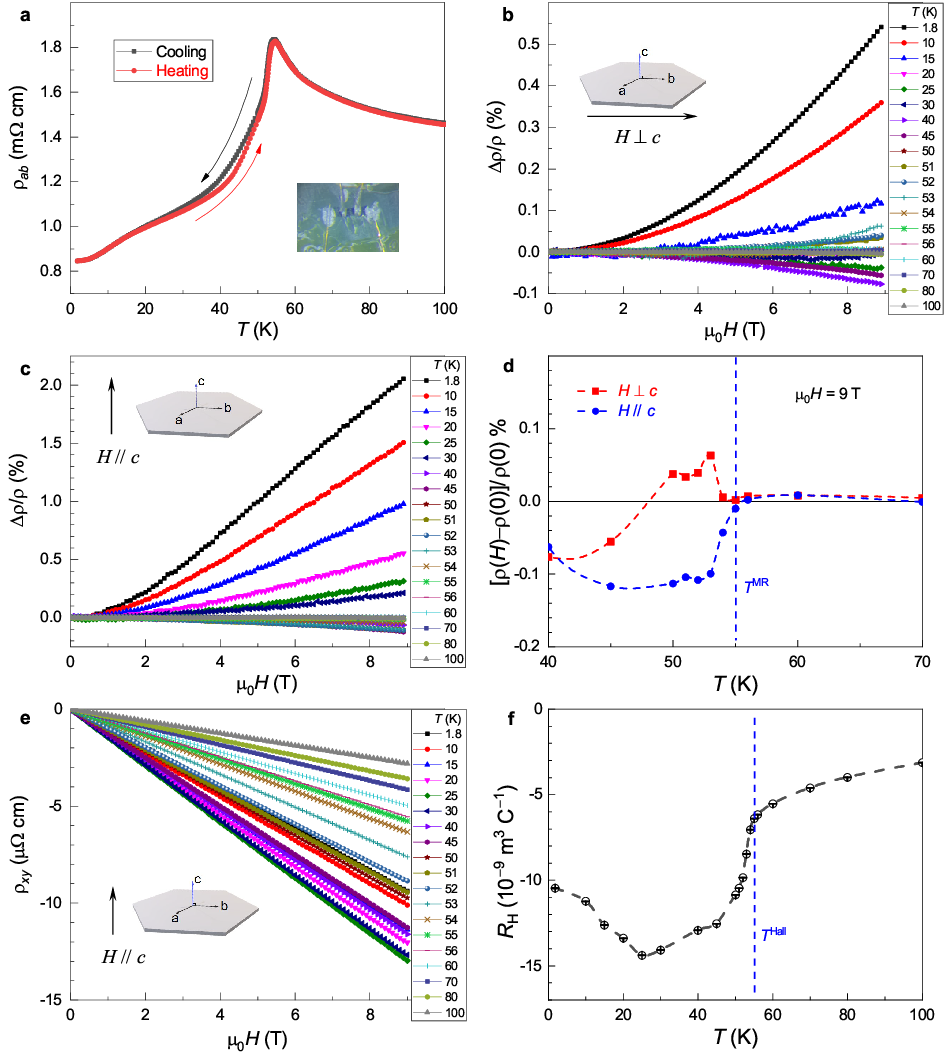}
\caption{\textbf{Extended Data Fig. 2 Electrical transport properties of \CCS~ crystals}. \textbf{a}, $\rho_{ab}(T)$ data in heating and cooling modes. The inset shows the four-electrode measurement configuration. \textbf{b,c}, Magneto-resistivity as functions of field parallel (\textbf{b}) and perpendicular to (\textbf{c}) the crystallographic $c$ axis. \textbf{d}, Magnetoresistance ($[\rho(H)-\rho(0)]/\rho(0)\%$) versus $T$. \textbf{e}, Hall resistivity as a function of magnetic fields at various temperatures. \textbf{f}, Temperature dependence of Hall coefficient, $R_\mathrm{H}$. Anomalies at $T^{\mathrm{MR}}\approx T^{\mathrm{Hall}}\approx$ 55 K are marked by the dashed vertical lines in \textbf{d,f}.}
\label{fig:Ex_PP}
\end{figure*}

Notably, there appears a resistivity peak at $T_\rho=$ 55 K. The corresponding anomaly is also manifested by a peak in the specific heat $C(T)$ (Fig.~\ref{fig:pp}c), a drop in the magnetic susceptibility $\chi(T)$ (Figs.~\ref{fig:pp}d,e), and changes in the magnetoresistance and Hall coefficient (Figs.~\ref{fig:Ex_PP}b-f in the Extended Data). No obvious thermal hysteresis was observed at around 55 K. Putting all these observations together, one concludes a second-order or weakly first-order phase transition for \CCS.

The $\chi(T)$ data at high temperatures exhibit a Curie-Weiss (CW) behaviour (Figs.~\ref{fig:pp}d,e), in contrast with the temperature-independent character in \CVS (refs.\cite{2020PRL.Cs135SC,2022PRM.LY}). We thus fitted the data with the formula $\chi(T)=\chi_0+C/(T+\theta_\mathrm{CW})$, where
$C$ refers to Curie constant, and $\theta_\mathrm{CW}$ is paramagnetic CW temperature.
To avoid unreliable results due to the mutual dependence among the three parameters, we carried out the data fitting with the constraint assuming the same value of Curie constant for the two field directions. Consequently, the best fit in the range of 150 K $<T<$ 300 K yields $C = 0.60\pm 0.11$ emu K mol$^{-1}$, from which an effective magnetic moment of $\mu_{\mathrm{eff}}= 1.26\pm0.12~\mu{_\mathrm{B}}$/Cr is obtained. The result 
suggests existence of Cr local moment at high temperatures. The temperature-independent term is fitted to be $\chi_0= 8.8\pm1.0\times10^{-4}$ ( $9.3\pm0.9\times10^{-4}$) emu mol$^{-1}$, and the fitted $\theta_{\mathrm{CW}}$ value turns out to be $314\pm50~ (385\pm56)$ K for $H ~\bot~ c$ ($H\parallel c$). Hence the Pauli-paramagnetic susceptibility is estimated to be $\sim1.3\times10^{-3}$ emu mol$^{-1}$ with the consideration of atomic core diamagnetism.
The large positive value of $\theta_{\mathrm{CW}}$ indicates strong AFM interactions between the Cr magnetic moments, which explains robustness of the phase transition against external magnetic fields (bottom inset of Fig.~\ref{fig:pp}c). Note that the magnetic frustration index\cite{1994.Ramirez}, $f=|\theta_{\mathrm{CW}}|/T_{\mathrm{N}}\approx$ 6-7, is moderately large, suggesting significant magnetic frustrations that commonly exist in a magnetic kagome lattice.

The nuclear magnetic resonance (NMR) measurement gives evidence of an antiferromagnetic (AFM) ordering (Fig.~\ref{fig:pp}f and Fig.~\ref{fig:Ex_NMR} in the Extended Data). The intensity of $^{123}$Sb-NMR spectra decreases abruptly below $T_{\mathrm{N}}\sim$ 55 K, signaling a shift in spectral weight, which is a typical feature of magnetic transitions\cite{2008JPSJ.Ba122.NMR}. At 2 K, a significant change in NMR spectra was observed: from one narrow line to two broad peaks (inset of Fig.~\ref{fig:pp}g). This splitting of NMR line strongly suggests an AFM order for the Cr spins because, in general, an AFM alignment of spins generates opposite internal magnetic fields at the Sb2 site. The severe spectral broadening dictates a large distribution of the internal magnetic field, implying that the magnetic structure should be somewhat complex. Indeed, a recent theoretical study\cite{2023arxiv.CC} suggests an altermagnetic order as the ground state for \CCS. One also notes that the Knight shift at 70 K is higher than that at 2 K (inset of Fig.~\ref{fig:pp}f), indicating that the spin susceptibility actually decreases with decreasing temperature. This result tells us that the observed magnetic susceptibility tail at low temperatures (Figs.~\ref{fig:pp}d,e) is of extrinsic origin, possibly due to tiny paramagnetic impurities and/or lattice imperfections.

\section{Structural modulation}

\begin{figure*}
\includegraphics[width=16cm]{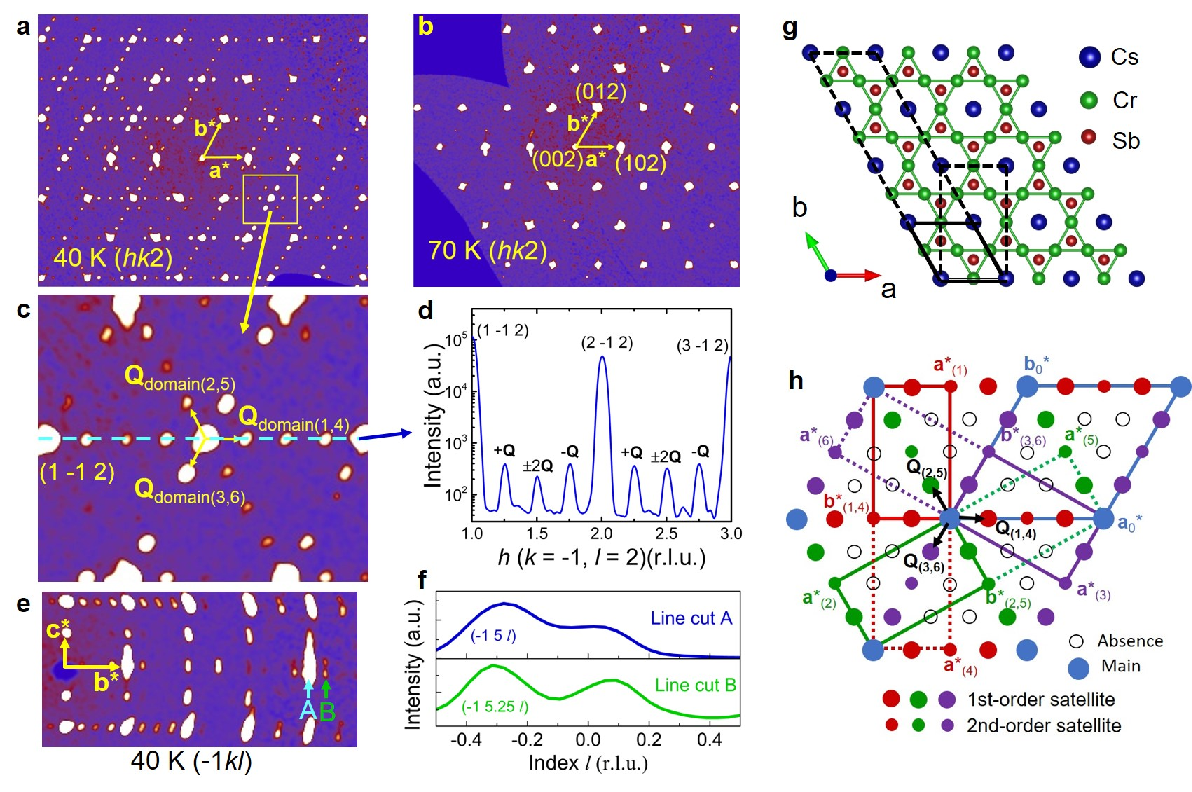}
\caption{$|$ \textbf{Structural modulations in CsCr$_3$Sb$_5$.} \textbf{a,b}, Reconstructed ($hk$2) planes of reflections at 40 and 70 K, respectively, with unit vectors $\mathbf{a}^*$ and $\mathbf{b}^*$ marked. \textbf{c}, A close-up around the main reflection (2$\bar{1}$2) in \textbf{a}, highlighting the satellite reflections that are indexed by one single $\mathbf{Q}$-vector with six twin domains. \textbf{d}, Cut along the blue dashed line marked in \textbf{c}. \textbf{e}, Reconstructed ($\bar{1}kl$) plane of reflections at 40 K. \textbf{f}, Line cuts along A and B marked in \textbf{e}. \textbf{g}, Crystal structure of CsCr$_3$Sb$_5$ viewed along the $c$ axis. The original hexagonal unit cell is marked by solid lines. The $C$-centered monoclinic and $a_{0}\times4a_{0}$ unit cells are marked with dashed lines. \textbf{h}, Schematic illustrations of the six monoclinic pseudo-orthohexagonal twin domains in the reciprocal lattice space projected along [001] direction. $\mathbf{a_\mathrm{0}^*}$ and $\mathbf{b_\mathrm{0}^*}$ are the original unit cell. $\mathbf{a}^*_{(i)}$ and $\mathbf{b}^*_{(i)}$ ($i = 1-6$) represent the lattice units of six individual monoclinic twin domains.
Domains 4-6 are marked with dashed lines for unit cells. $\mathbf{b}^*_{(1)}$, $\mathbf{b}^*_{(2)}$ and $\mathbf{b}^*_{(3)}$ share the same axis with $\mathbf{b}^*_{(4)}$, $\mathbf{b}^*_{(5)}$ and $\mathbf{b}^*_{(6)}$ in the projected view along $\mathbf{c}^*$, respectively.
Domains 1, 2 and 3 are connected by the three-fold rotation along $\mathbf{c}^*$. Domains 1 and 4 (the same for other pairs of domains) are correlated by a two-fold rotation along $\mathbf{b}^*$ due to the monoclinic distortion. The filled red, green and purple circles refer to satellite reflections attributed to $\mathbf{Q_{(1,4)}}$,$\mathbf{Q_{(2,5)}}$ and $\mathbf{Q_{(3,6)}}$, respectively. The satellite reflections of domain 1 and 4 are overlapped. The empty circles denote the absence of mixed-order satellite reflections between any two $\mathbf{Q_{i}}$ vectors in a hexagonal unit cell.}
	\label{CDW}
\end{figure*}

\begin{figure*}
\centering
\includegraphics[width=14cm]{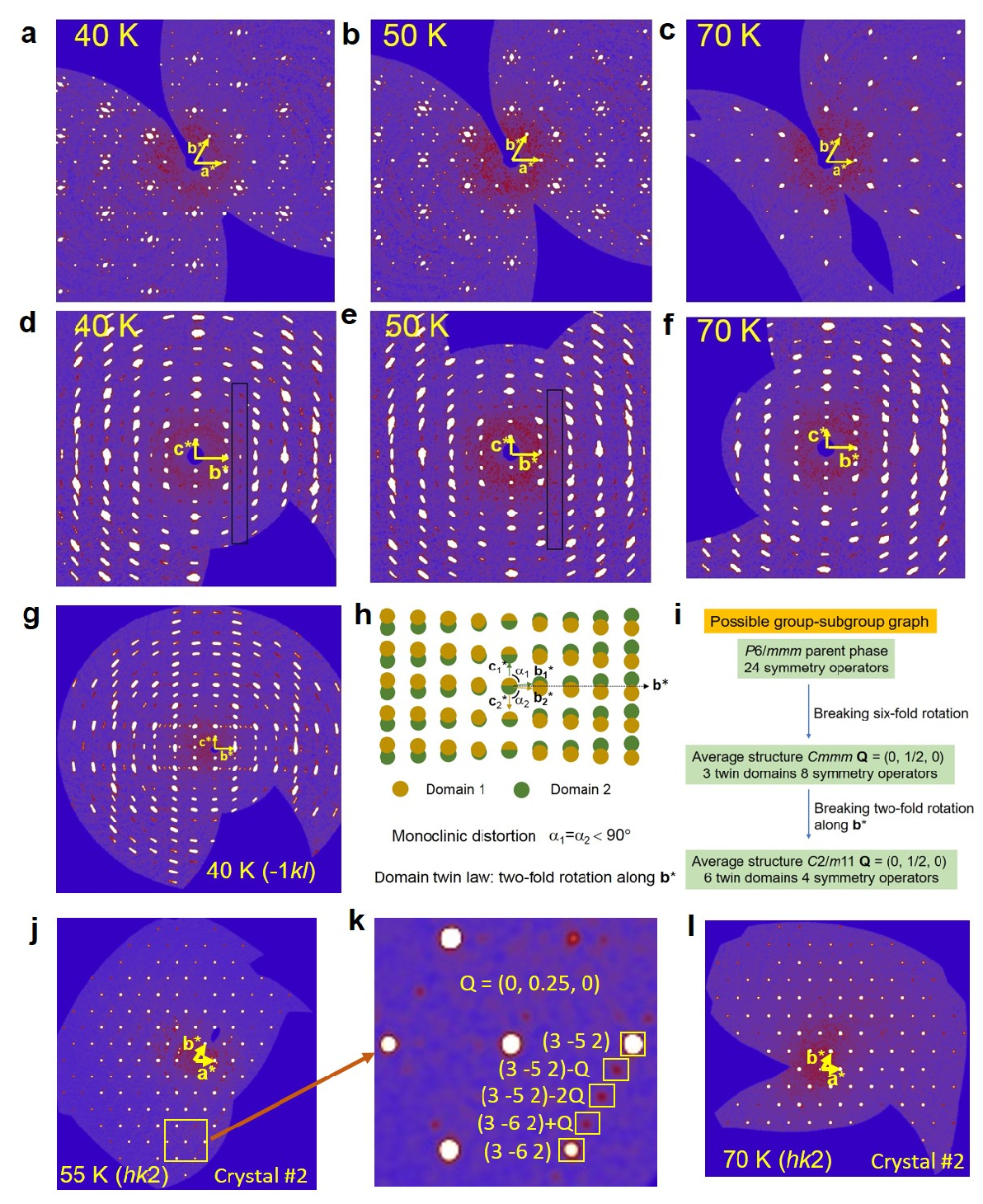}
\caption{\textbf{Extended Data Fig. 4 More information on the structural transition and modulations in CsCr$_3$Sb$_5$.} \textbf{a-c}, Reconstructed ($hk$0) planes of reflections at 40, 50, and 70 K, respectively, with unit vectors $\mathbf{a}^*$ and $\mathbf{b}^*$ marked. \textbf{d-f}, Reconstructed ($0kl$) planes of reflections at 40, 50, and 70 K, respectively. \textbf{g}, Reconstructed ($\bar{1}kl$) plane of reflections up to the highest resolution achieved by the data set. \textbf{h}, Illustration of monoclinic distortion in the reciprocal space to interpret the observed diffraction pattern in \textbf{g}. \textbf{i} Possible group-subgroup graph with the number of twin domains based on the qualitative analysis on satellite and main reflections for the structural modulation at 40 K. \textbf{j,l}, ($hk$2) planes of XRD reflections at 55 and 70 K, respectively, for another piece of the crystal. \textbf{k}, A close-up of the marked area in \textbf{j}.}
\label{fig:Ex_LTXRD}
\end{figure*}

To understand the phase transition in \CCS, we performed the single-crystal XRD down to 40 K. As shown in Figs.~\ref{CDW}a,b and Figs.~\ref{fig:Ex_LTXRD}a-c in the Extended Data, satellite reflections appear below $\sim$55 K, in addition to the primary Bragg diffractions in the $\mathbf{a^*b^*}$ plane. At first sight, these satellite spots seem to be related to a symmetry-equivalent triple-$\mathbf{Q}$ modulation vector that corresponds to a $4\times4$ superlattice based on the original hexagonal lattice. However, the intensities of the symmetry-related satellite spots by the six-fold rotation
are significantly unequal.
Therefore, the six-fold rotation symmetry of the hexagonal lattice is broken, and the diffraction pattern should be interpreted in terms of a multi-domain modulated structure with a single $\mathbf{Q}$ vector (1/4, 0, 0) based on a pseudo-hexagonal lattice. Such a single-$\mathbf{Q}$ modulation is indeed observed at 55 K (Figs.~\ref{fig:Ex_LTXRD}j,k in the Extended Data). In addition, peak splittings are observed especially at high diffraction angles in the $(\bar{1}kl)$ plane (Fig.~\ref{CDW}e), as clearly demonstrated by doing line cuts along $\mathbf{c}^*$ (Fig.~\ref{CDW}f). This points to a two-fold rotation symmetry breaking along $\mathbf{b}^*$ (Figs.~\ref{fig:Ex_LTXRD}g,h in the Extended Data).  
Notably, some mixed-order satellite spots between any two $\mathbf{Q}_{i}$ vectors are absent, indicating that these $\mathbf{Q}_{i}$ vectors are actually independent, coming from different domains. In the $\mathbf{b^*c^*}$ plane, nevertheless, no additional satellite spots along $\mathbf{c}^*$ direction are detected (Fig.~\ref{CDW}e and Figs.~\ref{fig:Ex_LTXRD}d-g in the Extended Data), in conformity with the $\theta-2\theta$ scan result shown in Fig.~\ref{fig:Ex_XRD}d of the Extended Data.
Combined with all the results above and following a group-subgroup law in space groups (Fig.~\ref{fig:Ex_LTXRD}i in the Extended Data), the low-temperature diffracting pattern can be equivalently interpreted by a modulated structure with a single $\mathbf{Q}$ vector of (0, 0.5, 0) based on the monoclinic lattice $a = a_0$, $b \approx \sqrt{3}a_0$, $c=c_0$, and $\alpha>90^\circ$.

As conduction electrons generally couple with the underlying lattice for a metallic system, the structural modulations point to a CDW-like instability\cite{2017AP.CDW}. Meanwhile, the magnetic susceptibility and NMR data above strongly suggest an AFM ordering at the phase transition. Therefore, it is reasonable to conjecture a concurrent intertwined DW order for the ground state, in which charge, spin, and lattice degrees of freedom couple together, something like the cases in metal Cr~\cite{2022NC.Cr} and the trilayer nickelate\cite{2020NC.4310Ni}.

\section{Superconductivity under pressure}

\begin{figure*}
\includegraphics[width=16cm]{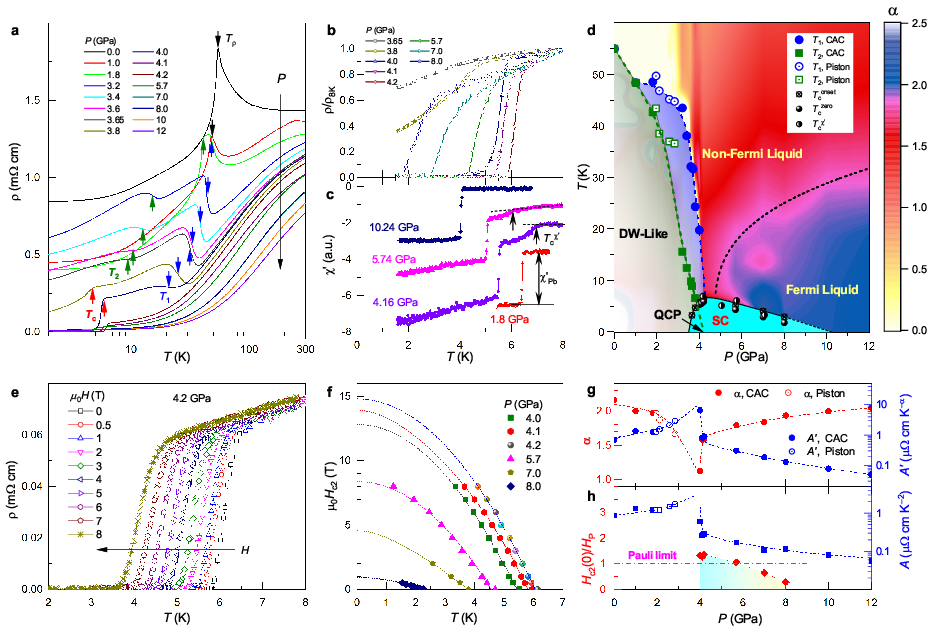}
\caption{\textbf{Superconductivity emerged from density-wave-like orders in \CCS.} \textbf{a}, $\rho(T)$ curves under high pressures. The blue, olive, and red arrows mark CDW-like, SDW-like, and superconducting transitions at $T_1$, $T_2$, and $T_\mathrm{c}$, respectively. \textbf{b}, Superconducting transitions at different pressures. \textbf{c}, Temperature dependence of ac susceptibility, $\chi'$, under high pressures. A piece of superconducting Pb was placed together with the sample as a reference material. \textbf{d}, The electronic $P-T$ phase diagram. DW, QCP, and SC denote density-wave, quantum critical point, and superconductivity, respectively. \textbf{e}, Superconducting transitions under various magnetic fields at 4.2 GPa. \textbf{f}, Upper critical fields as functions of temperature. \textbf{g}, Power $\alpha$ (left axis) and the coefficient $A'$ (right axis) as functions of pressure (see the text for details). \textbf{h}, The relative upper critical field to the Pauli-limited field (left axis) and the coefficient $A$ of the $T$-square term in $\rho=\rho_0+AT^2$ (right axis) as functions of pressure.
}
\label{fig:HP}
\end{figure*}

\begin{figure*}
\includegraphics[width=14cm]{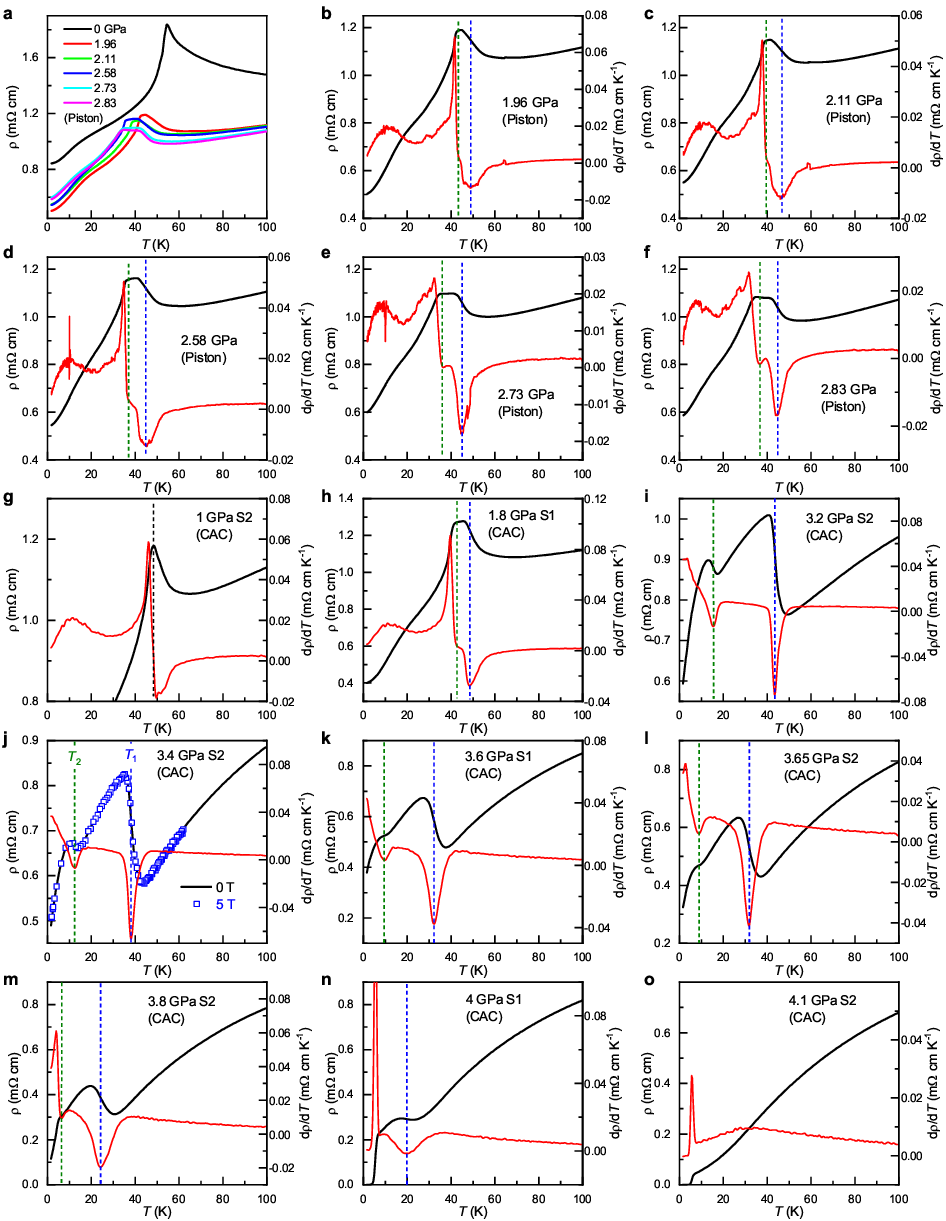}
\centering
\caption{\textbf{Extended Data Fig. 5 Close-ups of high-pressure $\rho(T)$ data for CsCr$_3$Sb$_5$ crystals.} \textbf{a-f}, $\rho(T)$ data obtained with a piston-type high-pressure cell. \textbf{g-o}, $\rho(T)$ data measured using a cubic anvil cell (CAC). Evolution of the two characteristic temperatures of $T_1$ and $T_2$ (labelled in \textbf{j}), marked by the blue and olive dashed lines respectively, can be tracked.}
\label{fig:Ex_HP1}
\end{figure*}

\begin{figure*}
\includegraphics[width=12cm]{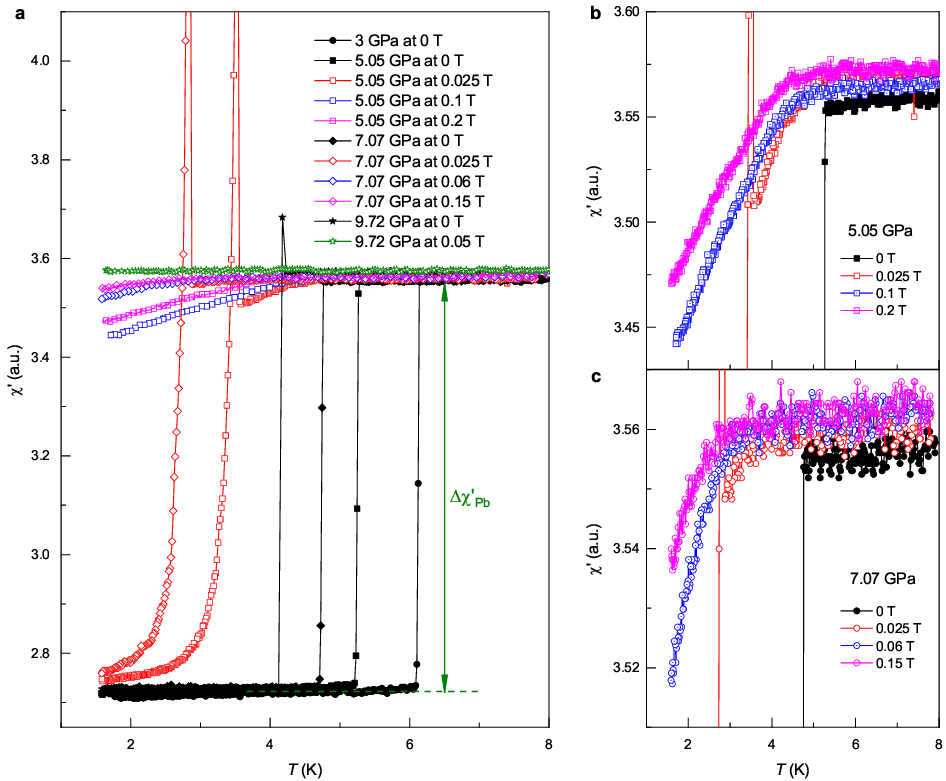}
\centering
\caption{\textbf{Extended Data Fig. 6 Additional high-pressure magnetic susceptibility measurements for CsCr$_3$Sb$_5$ crystals.} \textbf{a}, Temperature dependence of ac susceptibility $\chi'$ at zero field as well as under small magnetic fields for suppressing superconductivity of the reference material Pb. \textbf{b,c}, Close-ups of \textbf{a} highlighting the superconducting transitions at 5.05 GPa (\textbf{b}) and 7.07 GPa (\textbf{c}).}
\label{fig:Ex_HP2}
\end{figure*}

\begin{figure*}
\includegraphics[width=16cm]{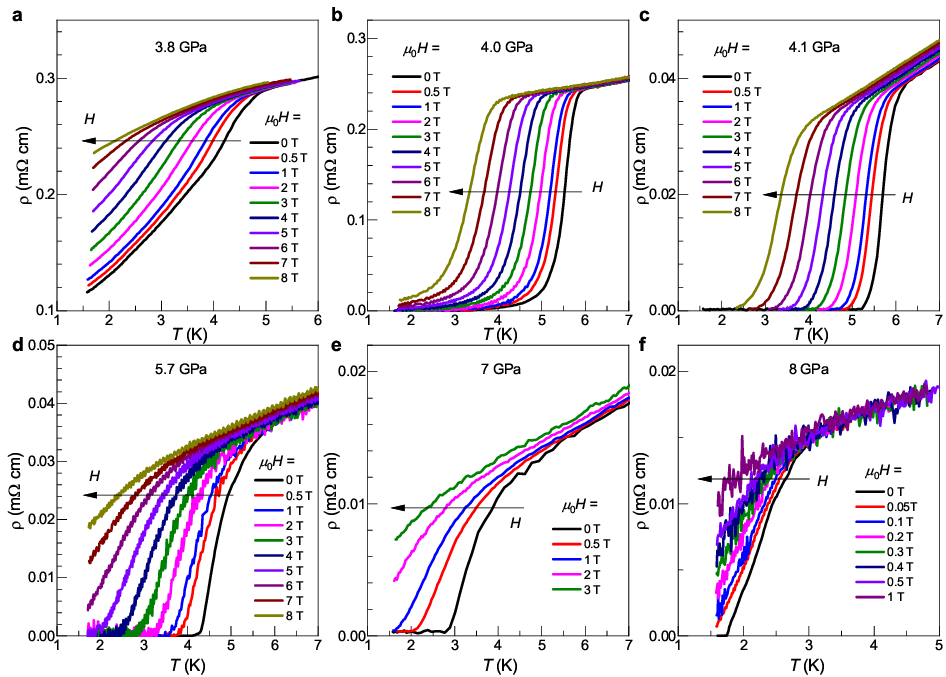}
\centering
\caption{\textbf{Extended Data Fig. 7 Additional high-pressure $\rho(T)$ data under magnetic fields for CsCr$_3$Sb$_5$ crystals.} \textbf{a-f}, Plots of $\rho(T)$ at $P=$ 3.8, 4.0, 4.1, 5.7, 7.0, and 8.0 GPa, respectively, from which the upper critical fields at zero temperature were extracted.}
\label{fig:Ex_HP3}
\end{figure*}

Figure~\ref{fig:HP}a shows the $\rho(T)$ data of \CCS~ under various pressures up to 12 GPa. One sees that the resistivity decreases monotonically with pressure, and the metallicity is steadily enhanced. At $P>$ 1 GPa, the resistivity peak evolves into two anomalies at $T_1$ and $T_2$, respectively (for the detailed evolution see Fig.~\ref{fig:Ex_HP1} in the Extended Data).
There is a remarkable resistivity jump at $T_1$, implying a CDW ordering because the latter generally opens an energy gap. Comparatively, the anomaly at $T_2$ is rather mild, which is putatively attributed to an SDW ordering. If so, the structural modulation/distortion expected at $T_1$ will release the geometric frustration, which helps the AFM ordering at $T_2$.
With increasing pressure, both $T_1$ and $T_2$ decrease monotonically, and superconductivity gradually emerges for $P>$ 3.6 GPa. At 3.65 and 3.8 GPa, both superconducting and DW-like transitions are observed (Figs.~\ref{fig:Ex_HP1}l,m and Fig.~\ref{fig:Ex_HP3}a in the Extended Data), suggesting coexistence of superconductivity and DW-like orders.
At 4 GPa $\leq P\leq$ 8 GPa, the superconducting transitions are clearly seen in Fig.~\ref{fig:HP}b with zero resistance, and bulk superconductivity is confirmed by the ac magnetic susceptibility ($\chi'$) measurement (Fig.~\ref{fig:HP}c and Figs.~\ref{fig:Ex_HP2}a-c in the Extended Data).
The highest superconducting transition temperature of $T_\mathrm{c}^{\mathrm{onset}}=$ 6.4 K is achieved at 4.2 GPa. 
For $P\geq$ 10 GPa, no superconductivity can be observed down to 1.6 K.

Figure~\ref{fig:HP}e shows the superconducting transitions under different magnetic fields at 4.2 GPa (for data of other pressures see Figs.~\ref{fig:Ex_HP2}a-f in the Extended Data). As expected, $T_\mathrm{c}$ shifts to lower temperatures with increasing fields.
Here we used the criteria of 50\% of normal-state resistivity for determining $T_\mathrm{c}(H)$, from which the upper critical fields $\mu_0H_{\mathrm{c2}}(T)$ are derived (Fig.~\ref{fig:HP}f). The $\mu_0H_{\mathrm{c2}}(T)$ data can be well described by the Ginzburg-Landau equation, $H_{\mathrm{c2}}(T) = H_{\mathrm{c2}}(0)[1-(T/T_c)^2]$. The fitted zero-temperature $\mu_0H_{\mathrm{c2}}(0)$ are 11.95 and 14.34 T for $P=$ 4.0 and 4.2 GPa, respectively, which exceeds the Pauli limit of $\mu_0H_{\mathrm{P}}\sim 1.84 T_\mathrm{c}$ (in tesla). Note that field direction was not sure for the present high-pressure measurement, and the anisotropy of $H_{\mathrm{c2}}(T)$ is left for future study.

With the above high-pressure results, the electronic $P-T$ phase diagram for \CCS~is established (Fig.~\ref{fig:HP}d). At ambient pressure, there is a charge- and spin-DW-like ordering at 55 K. When the applied pressure exceeds $\sim$1 GPa, it evolves into two successive transitions, probably associated with CDW and SDW orderings, respectively. Both the DW-like transition temperatures, $T_1$ and $T_2$, are suppressed gradually before going to absolute zero. The critical pressure at $T_2\rightarrow$ 0 is extrapolated to be $P_\mathrm{c}\approx$ 4.2 GPa where $T_\mathrm{c}$ and $H_{\mathrm{c2}}(0)$ achieve their maximal values, respectively. In addition, there is a narrow region of 3.6 GPa $< P <$ 4.1 GPa, in which superconductivity likely coexists with the DW-like orders. All these results resemble those of iron-based superconductors\cite{2014review.Matsuda,2009PRB.Ba122.Canfield}, CrAs\cite{2014NC.CrAs}, and MnP\cite{2015PRL.MnP}, pointing to quantum criticality in the present system.

\begin{figure*}
\includegraphics[width=16cm]{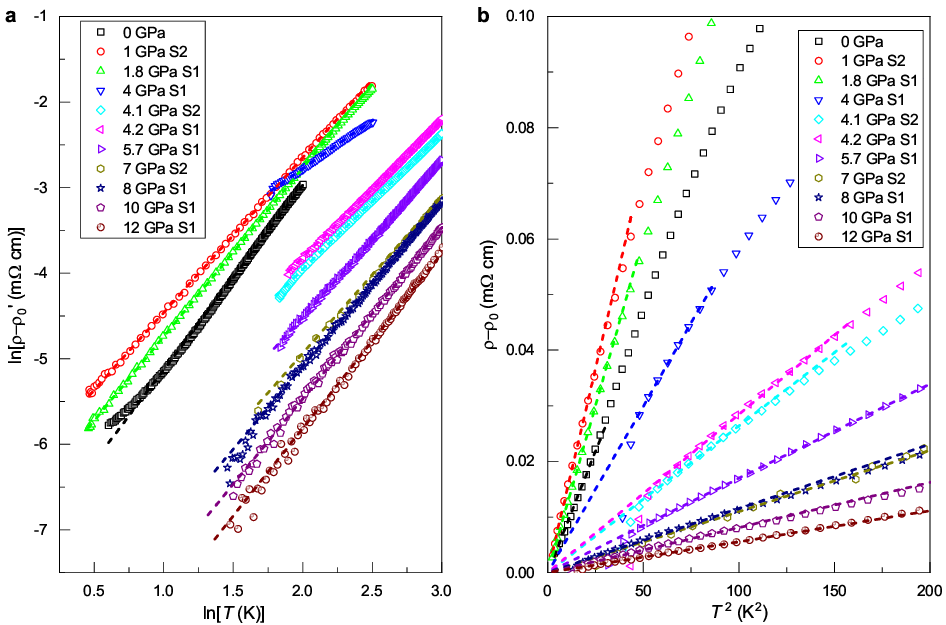}
\centering
\caption{\textbf{Extended Data Fig. 8 Temperature dependent normal-state resistivity under high pressures for CsCr$_3$Sb$_5$ crystals.} \textbf{a}, Bilogarithmic graph for the $\rho(T)$ relations, which gives the power $\alpha$ in the formula $\rho=\rho_0'+A'T^{\alpha}$. \textbf{b}, Plot of ($\rho-\rho_0$) versus $T^2$, which yield the coefficient $A$ and $\rho_0$ as described in the main text.}
\label{fig:Ex_HP4}
\end{figure*}

One of the most important hallmarks of quantum criticality is a non-Fermi-liquid (or, strange-metal) behaviour, which is indeed observed from the normal-state resistivity just above $T_\mathrm{c}$. The data fitting (Fig.~\ref{fig:Ex_HP4}a in the Extended Data) with the power law $\rho = \rho_0'+A'T^{\alpha}$ shows that,
at 4 GPa, $\alpha$ goes to $\sim$1.0, distinct from $\sim$2.0 at the lower and higher pressures (Fig.~\ref{fig:HP}g), suggesting the breakdown of Fermi-liquid state at around the QCP (ref.\cite{2007RMP.QPT}). The detailed variations of $\alpha$ are also displayed in the coloured contour plot, in which the non-Fermi-liquid regime is marked.
If assuming Fermi-liquid scenario in the low-temperature limit, alternatively, the data fitting with the formula $\rho=\rho_0+AT^2$ (Fig.~\ref{fig:Ex_HP4}b in the Extended Data) yields a remarkable enhancement of the coefficient $A$ (Fig.~\ref{fig:HP}h) at $P\rightarrow$ 4 GPa. A similar divergence behaviour is seen for the parameter $A'$ (Fig.~\ref{fig:HP}g). This result further corroborates a QCP at around 4 GPa. In addition, the large value of $A$ actually suggests USC in pressurized \CCS, because the $T_\mathrm{c}/T_\mathrm{F}$ ($T_\mathrm{F}$ is the Fermi temperature) values estimated just lie in the USC region covering cuprates, iron-based pnictides/chalcogenides, and heavy-fermion materials (Fig.~\ref{fig:Ex_USC} in the Extended Data). 

\section{DFT Calculations}

\begin{figure*}
\includegraphics[width=16cm]{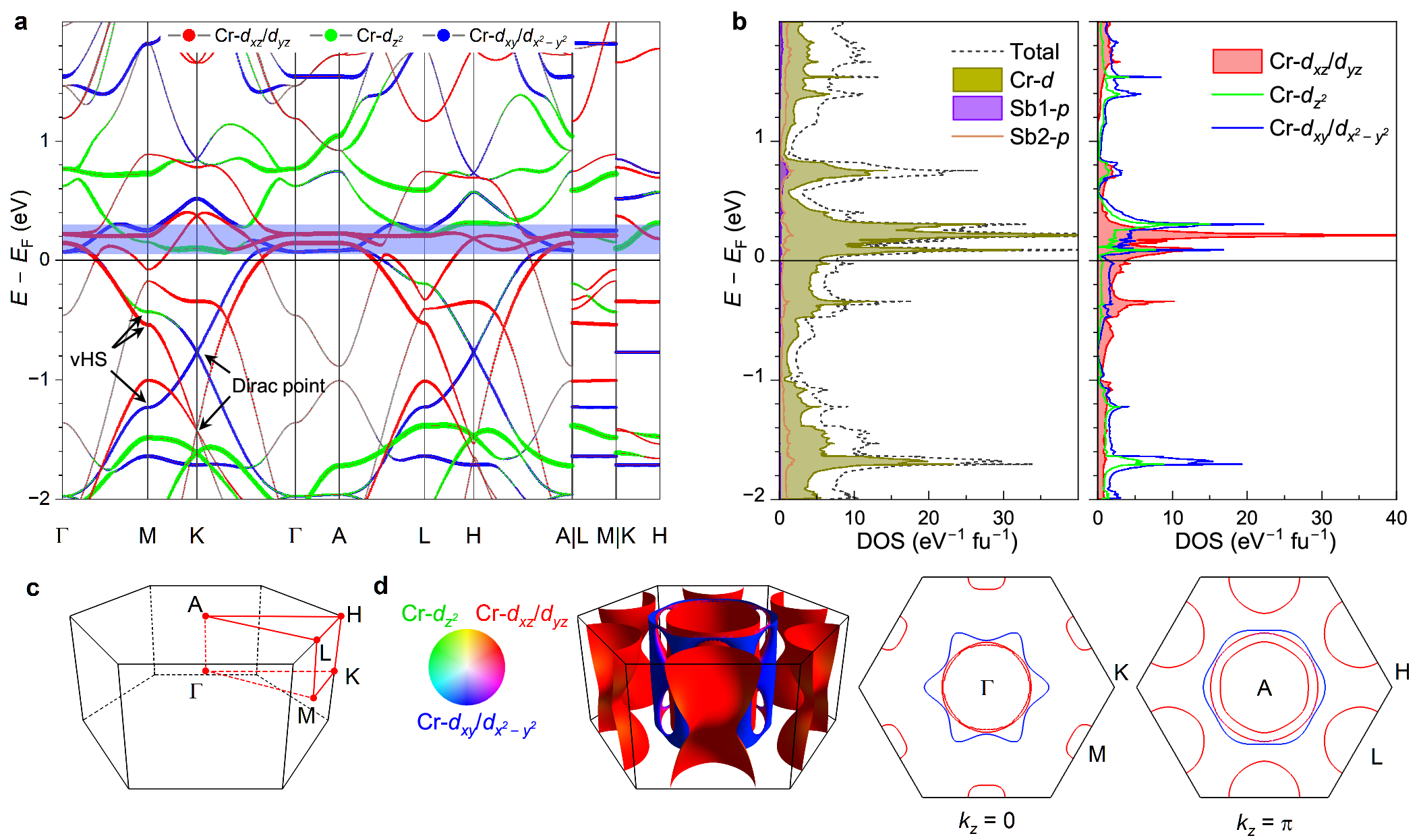}
\caption{\textbf{Electronic structure of \CCS~ by the DFT calculations without spin-orbit coupling.} \textbf{a}, The band structure highlighting contributions from Cr-$d_{xz}/d_{yz}$, Cr-$d_{xy}/d_{x^2-y^2}$, and Cr-$d_{z^2}$ orbitals. The van Hove singularity (vHS) and the Dirac point are indicated by the arrows, and the flat bands are marked by the transparent blue stripe. \textbf{b}, Density of states (DOS) contributed from each atom (left) and different Cr-3$d$ orbitals (right). \textbf{c}, The Brillouin zone with high-symmetry points marked. \textbf{d}, The merged Fermi-surface (FS) sheets (left), FS slices at $k_z$ = 0 (middle) and $\pi$ (right). The red, blue, and green colours denote the Cr-$d_{xz}/d_{yz}$, Cr-$d_{xy}/d_{x^2-y^2}$, and Cr-$d_{z^2}$ components, respectively.}
\label{fig:BS}
\end{figure*}

To understand the basic electronic structure of CsCr$_3$Sb$_5$, we performed the first-principles density-functional-theory (DFT) calculations for the high-temperature hexagonal phase. The calculated band structure (Fig.~\ref{fig:BS}a), which highlights the components of distinct Cr-3$d$ orbitals, shows a metallic behaviour with three bands crossing the Fermi level, $E_\mathrm{F}$.
As a result, two hole-type Fermi surfaces (FSs) around the $\Gamma$A line and one electron-type FS around the ML line are derived (Fig.~\ref{fig:BS}d). All the FS sheets are quasi-two dimensional, which explains the large anisotropy of resistivity.
For $d_{xz/yz}$ and $d_{xy/x^2-y^2}$ derived bands, notably, they show characteristic kagome-like electronic structures with van Hove singularities, Dirac points, and flat bands. Those flat bands, which are marked with the blue stripe, are somewhat distorted along certain directions owing to the anisotropic hoppings between different 3$d$ orbitals and the hybridization between Cr-kagome and Sb2-honeycomb lattices.

The Fermi level in \CCS is only $\sim$0.1 eV lower below the nearest flat band, meanwhile it is $\sim$0.5 eV higher above the van Hove points, in contrast to \CVS~(Fig.~\ref{fig:Ex_DFT}c in the Extended Data). This is primarily due to the different electron filling in the Cr- or V-3$d$ orbitals and, consequently, the FS topology is very distinct from that of \CVS~(refs.\cite{2020PRL.Cs135SC,2021PRX.Ortiz}). This implies that the CDW-like transition in \CCS~may not be driven by FS nesting and, instead, electron-phonon interactions and electron correlations might play an important role.
Under a pressure of 5 GPa, interestingly, the flat bands move up
with valence bandwidths moderately broadened
(Fig.~\ref{fig:Ex_DFT}d in the Extended Data).
As a consequence, the band edges around the M point shift up, the system undergoes a Lifshitz transition\cite{1960SPJETP.Lifshitz}, where the quasi-two dimensional FS around the ML line breaks into a large electron pocket (around the L point) and a tiny hole pocket (around the M point). Such a unique feature in electronic structure might be in relation with the suppression of DW-like orders with applying pressure.

The electronic states at around $E_\mathrm{F}$ are mostly contributed from the Cr-3$d$ orbitals (Fig.~\ref{fig:BS}b). Among them, Cr-3$d_{xz/yz}$ and
Cr-3$d_{xy/x^2-y^2}$ dominate the states at $E_\mathrm{F}$.
The total density-of-states at $E_\mathrm{F}$ is $D(E_\mathrm{F})=$ 7.7 states eV$^{-1}$ fu$^{-1}$. This gives a theoretical value of the Pauli-paramagnetic susceptibility, $\chi_\mathrm{P}^{T}=\mu_\mathrm{B}^2D(E_\mathrm{F})=$ $2.5\times10^{-4}$ emu mol$^{-1}$, about 1/5 of the experimental value estimated above. Meanwhile, the bare $D(E_\mathrm{F})$ value corresponds to an electronic specific-heat coefficient of $\gamma_{0}=\frac{1}{3}\pi^{2}k_{\mathrm{B}}^2D(E_{\mathrm{F}})=$ 18.1 mJ K$^{-2}$ mol$^{-1}$, which is only $\sim$1/6 of the experimental value. Note that the bare $D(E_\mathrm{F})$ should be lower due to the DW-like order. This means that the electron-mass renormalization factor could be even larger due to strong correlation effect.

\section{Concluding Remarks}
Above we demonstrate that CsCr$_3$Sb$_5$ bears contrastingly different properties in comparison with its structurally analogous compound  CsV$_3$Sb$_5$ (Table 2 in the Extended Data). Here we make a summary as follows:
i) CsCr$_3$Sb$_5$ behaves as a correlated metal with a large electron-mass renormalization and significant magnetic frustration. It undergoes a phase transition at 55 K associated with AFM DW-like ordering at ambient pressure. 
ii) The DW-like order
has a single-$\mathbf{Q}$ modulation with a $4\times1\times1$ supercell based on a pseudo-hexagonal lattice.
Owing to the dynamically unstable lattice\cite{2023arxiv.CC} and the coupling between spin and charge degrees of freedoms, electron-phonon and/or magnetic interactions may play an important role for the DW-like order.

iii) Under high pressure, the DW-like orders are gradually suppressed, and a superconducting dome emerges nearby a QCP at about 4.2 GPa. This is in contrast with the $A$V$_3$Sb$_5$ family, the latter of which exhibits superconductivity already at ambient pressure and, with applying pressure, superconductivity does not disappear\cite{2022nature.HP.CXH,2021PRB.HP-Cs135.YangZR,2021PRL.HP.ChengJG}.

\begin{figure*}
\includegraphics[width=14cm]{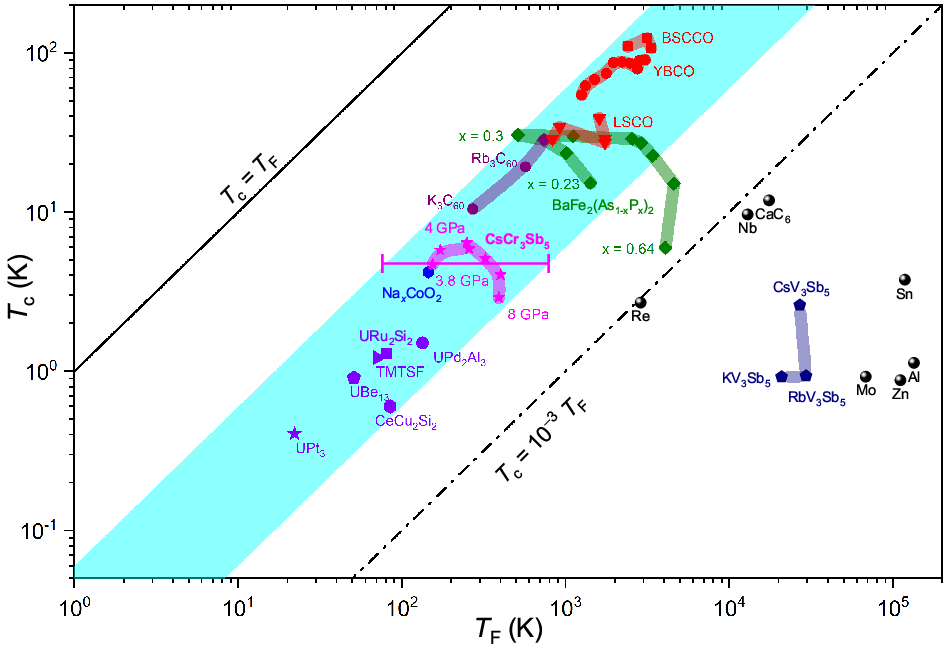}
\centering
\caption{\textbf{Extended Data Fig. 9 Superconducting critical temperature $T_\mathrm{c}$ versus Fermi temperature $T_\mathrm{F}$ for various superconductors.} The $T_\mathrm{F}$ values of CsCr$_3$Sb$_5$, CsV$_3$Sb$_5$, RbV$_3$Sb$_5$, and KV$_3$Sb$_5$ are estimated by the relation, $A\propto(m^*/m_b )^2/T_\mathrm{F}^2$, where $A$ is the coefficient of $T$-square resistivity and $m^*/m_b$ is the electron-mass renormalization factor [K. Behnia, Ann. Phys. 534, 2100588 (2022)]. The electron-mass renormalization is estimated by comparing the theoretical and experimental the electronic specific-heat coefficient, $m^*/m_b\approx \gamma_{\mathrm{exp}}/\gamma_{\mathrm{bare}}$, where $\gamma_{\mathrm{bare}}$ is obtained from the DFT calculations. At ambient pressure, the $\gamma_{\mathrm{exp}}/\gamma_{\mathrm{bare}}$ value for \CCS~is estimated to be 6, if the influence of DW-like order is roughly neglected. At the moderately high pressure, it generally tends to decrease to some extent because of pressure-induced band broadening. Here we adopt this ``upper limit'', and then the $T_\mathrm{F}$ values at 4-8 GPa for \CCS~are estimated to be 150-400 K. The uncertainties of $T_\mathrm{F}$ are not expected to exceed 100\%, as marked by the vertical bar. The coefficient $A$ of V-based kagome superconductors is obtained by fitting the low-$T$ data using the formula $\rho(T)=\rho_0+AT^2$ (original data from [CsV$_3$Sb$_5$: B. R. Ortiz, et al., Phys. Rev. Lett. 125, 247002 (2020); RbV$_3$Sb$_5$: Q. Yin, et al., Chin. Phys. Lett. 38, 037403 (2021). KV$_3$Sb$_5$: B. R. Ortiz, et al., Phys. Rev. Mater. 5, 034801 (2021)]). And the $m^*/m_b$ value for $A$V$_3$Sb$_5$ is about 1.0 (see the related data in Table 2). The $T_\mathrm{F}$ data of other superconductors are taken from the references [T. Shibauchi, A. Carrington, and Y. Matsuda, Annu. Rev. Condens. Matter Phys. 5, 113 (2014)] and [Y. Cao et al., Nature 556, 44 (2018)]. The highlighted stripe region, where CsCr$_3$Sb$_5$ is located, is almost exclusively occupied by known USCs.}
\label{fig:Ex_USC}
\end{figure*}

USC typically emerges from a spin- and/or charge-ordered state in a correlated electron system, as exemplified in cuprates, iron-based pnictides/chalcogenides, and heavy-fermion materials\cite{2013RMP.Scalopino,2017AP.Stewart}. Nevertheless, so far there is still no consensus on the mechanism of USC\cite{2011science.Norman}, although spin fluctuations are generally considered as a common pairing glue\cite{2013RMP.Scalopino}. Here CsCr$_3$Sb$_5$ bears similarities with the previous USCs in terms of electron correlations, evolution of superconductivity, intertwined orders, and the $T_\mathrm{c}/T_\mathrm{F}$ value in particular (Fig.~\ref{fig:Ex_USC} in the Extended Data). This possible realization of USC in the correlated kagome metal supplies a unique example, which may shed light on the mechanism of USC.

There are a lot of issues open for the future studies. For instance, further confirmation of USC in pressurized \CCS~is important. It is also appealing to reveal the superconducting pairing symmetry. For these purposes, experimental explorations for the ambient-pressure superconductivity via chemical doping/substitutions and/or gate-voltage regulation are highly deserved. The nature of the phase transition, especially the form of magnetic structure, as well as its evolution under pressure, is of special interest given its intimate relationship with the possible USC.

\noindent {\bfseries Acknowledgments} We thank Y.T. Song for the assistance in the low-temperature single-crystal X-ray diffraction measurements. This work is supported by the National Natural Science Foundation of China (grant nos. 12050003, 12004337, 12025408, 11921004, 11974405, 12204298, 12274364, and 12274369), the National Key R\&D Program of China (grant nos. 2022YFA1403202, 2021YFA1400200, 2023YFA1406100, and 2022YFA1403402), the Key R\&D Program of Zhejiang Province, China (grant no. 2021C01002), the Strategic Priority Research Program of CAS (grant nos. XDB33000000 and XDB33010100), and the Outstanding Member of Youth Promotion Association of CAS (grant no. Y2022004).
The high-pressure experiments and the NMR experiments were carried out at the Cubic Anvil Cell (CAC) and the High-field NMR stations of Synergic Extreme Condition User Facility (SECUF), respectively.

\noindent
{\bfseries Author Contributions} G.-H.C. coordinated the work, co-conceived the experiments with Y.L., and interpreted the result in discussion with J.-G.C., R.Z., J.-K.B., Y.L., X.-F.X, and C.C. The high-pressure experiments were performed by Z.-Y.L., P.-T.Y., and B.-S.W. under the leadership of J.-G.C. J.-K.B contributed the structural analysis with the help from J.-Y.L. The NMR measurement was done by Q.X.S., J.L., and J.Y., supervised by R.Z. The theoretical calculations were made by L.-W.J., S.-Q.W., C.-C.X., H.J., and C.C. Crystals were grown by Y.L., W.-L.C., J.-Y.L., and C.-C.L. The ambient-pressure physical-property measurements were done by Y.L., W.-Z.Y., Q.T., Z.R., and Z.-A.X. The paper was written by G.-H.C., J.-G.C., R.Z., J.-K.B., Y.L., and Z.-Y.L. All co-authors made comments on the manuscript.

\noindent
{\bfseries Competing interests} The authors declare no competing financial interests.

\noindent
{\bfseries Correspondence and requests for materials} should be addressed to Guang-Han Cao, Jin-Guang Cheng, and Rui Zhou.

\noindent
{\bfseries Data availability} The data shown in the main figures are provided in the Source data.

\vspace{12pt}

 \clearpage
 \newpage

 \section*{Methods}
\textbf{Crystals Growth and Characterizations}\\
Single crystals of CsCr$_3$Sb$_5$ were grown using a self-flux method from the constituent elements Cs (Alfa 99.999$\%$), Cr (Alfa 99.99$\%$), and Sb (Aladdin 99.999$\%$). Eutectic composition in the CsSb$-$CsSb$_2$ quasi-binary system was employed as the flux.
The mixture of Cs, Cr, and Sb was loaded into an alumina crucible, and then it was sealed in a Ta tube by arc welding under argon atmosphere. The Ta tube was protected from oxidation by sealing in an evacuated silica ampoule. The sample loaded assembly was heated in a furnace to 850-900~$^{\circ}$C holding for 18 h, and subsequently cooled slowly at a rate of 2-4~$^{\circ}$C/h to 500-600~$^{\circ}$C. Thin crystalline flakes can be found in the melts.
The harvested crystals are stable in air with size up to 0.5 $\times$ 0.5 $\times$ 0.02 mm$^3$.

Single-crystal XRD was carried on a Bruker D8 Venture diffractometer with Mo K$\alpha$ radiation. A piece of CsCr$_3$Sb$_5$ crystal with dimensions of $0.018\times0.168\times0.172$ mm$^{3}$ was mounted on the sample holder using the oil of polybutenes. A flow of cryogenic helium gas was used to cool the crystal from room temperature to 40 K first and then to warm up to 50 and 70 K successively. We also measured another piece of the crystal at 55, 58, and 70 K. A full data set was collected at each temperature. The data reduction including integration and scaling was done by the commercial software packages APEX4. The reconstructed images in the reciprocal space from the raw frames were produced using the reciprocal unit vectors of the hexagonal lattice by the software CrysAlis$^{\mathrm{Pro}}$ (CrysAlis Pro Version 171.40.53, Rigaku Oxford Diffraction.) The crystal structure of CsCr$_3$Sb$_5$ was initially solved by SUPERFLIP\cite{2007JACrst.SUPERFLIP}. and then refined against the structure factor $F$ in JANA2006\cite{2014.JANA}.
We also performed $\theta-2\theta$ scan on a PANalytical X-ray diffractometer (Model EMPYREAN) with a monochromatic CuK$\alpha_1$ radiation, 
which generates the $(00l)$ diffraction pattern.
The chemical composition of the as-grown crystals was determined using EDX spectroscopy on a scanning electron microscope (Hitachi S-3700N) equipped with Oxford Instruments X-Max spectrometer.

\noindent\textbf{Physical Property Measurements}\\
\noindent The measurements of electrical resistivity, magneto-resistivity, Hall effect, and specific heat were carried out on a physical property measurement system (PPMS-9, Quantum Design). The resistivity was measured by a standard four-terminal method using silver paste for making the electrodes. For the measurement of $\rho_{c}(T)$, the electrodes were made on both sides of the crystal, and the current electrodes were prominently larger than the voltage ones, such that the electric current flows basically homogeneously along the $c$ axis. The Hall coefficient and magnetoresistance with magnetic field parallel to $c$ axis were simultaneously measured on a nearly square shaped crystal with a six-electrode configuration. The resistance and Hall signals were obtained respectively by symmetrizing and antisymmetrizing the data collected in reversed magnetic fields. Specific heat was measured using thermal relaxation method with dozens of the crystals (total mass was 0.29~mg). The samples were glued on the heat capacity puck with N grease. The data of addenda were measured in advance. The measurements were carried out for three times at each temperature, and the final $C(T)$ data were obtained by averaging. The temperature-rise parameter was set to the minimal value of 1\% in the PPMS-9 for the best temperature resolution.

The magnetic measurements were performed on a magnetic property measurement system (MPMS-3, Quantum Design). Samples with mass of 0.17 mg were carefully amounted on the sample holder. For the measurements with field perpendicular to the $c$ axis, pieces of crystals were attached on the quartz paddle with a little N grease. For the case with field parallel to the $c$ axis, an additional high-purity quartz plate was used to hold the samples. The quartz plate was stuck to the quartz paddle with GE varnish. The assembly without samples was measured in advance as addenda.

The NMR measurements were performed with a phase-coherent pulsed NMR spectrometer. To get sufficient NMR signal intensity, the $^{121/123}$Sb-NMR spectra were measured on a collection of $\sim$20 single crystals ($\sim$0.1 mg) acquired by sweeping the frequency point by point and integrating the spin-echo signal. We stacked the single-crystal flakes along the $c$ direction, ensuring the applied magnetic field along the $c$ axis. Above $\mu_0H_0 = 16$ T, the NMR measurements were conducted by using the 26-T high-field NMR system in the SECUF at Huairou, Beijing.

\begin{figure*}
\centering
\includegraphics[width=12cm]{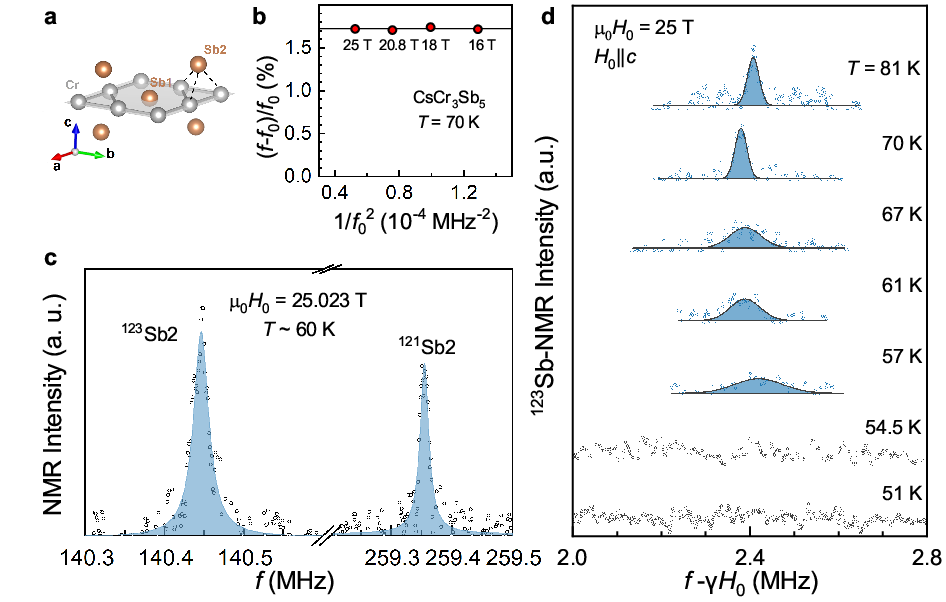}
\caption{\textbf{Extended Data Fig. 3 More information about Sb2 NMR spectra in \CCS.} \textbf{a}, Crystal structure of \CCS~ showing distinct Sb1 and Sb2 sites. \textbf{b}, Frequency dependence of the peak frequencies in \CCS~ at 70 K obtained at fixed fields marked. Lines are linear fits which indicate the asymmetry parameter $\eta$ is 0. \textbf{c}, $^{123}$Sb2 and $^{121}$Sb2 NMR lines at 60 K and under a magnetic field of $\mu_0H_0$ = 25 T. \textbf{d}, $^{123}$Sb-NMR spectra at various temperatures under a magnetic field of 25 T parallel to the $c$ axis. Solids lines are the Gaussian fit.}
\label{fig:Ex_NMR}
\end{figure*}

There are two types of Sb sites in CsCr$_3$Sb$_5$. Sb1 is located at the center of the chromium hexagon, while Sb2 is located above the chromium triangle, as displayed in Fig.~\ref{fig:Ex_NMR}a of the Extended Data. In principle, we should observe two NMR central lines. However, due to the small quantity of the sample used for NMR experiments, we had a poor signal-to-noise ratio, and only one line is visible in $^{121}$Sb- and $^{123}$Sb-NMR spectra as shown in Fig.~\ref{fig:Ex_NMR}c. Given that the atomic ratio between Sb1 and Sb2 is 1:4, it is most likely that the observed line is from the Sb2 site.

$^{123}$Sb nuclei ($I$ = 7/2) has a quadrupole moment $Q$ that couples to the electric field gradient. This interaction is responsible for the second-order quadrupolar shift of the central line that adds to the Knight shift $K$. The total frequency shift can be expressed as,
$\frac{f-f_0}{f_0}=K+\frac{\nu_{c}^{2}\eta^{2}}{12(1+K)f_0^2},$
where $f_0 = ^{123}\gamma H_0$ is the reference frequency, and $^{123}\gamma$ is the gyromagnetic ratio of $^{123}$Sb. $\nu_{c}$ is the component of the quadrupole coupling tensor along its main axis ($z$ along the crystallographic $c$ direction) and $\eta=|\frac{v_{XX}-v_{YY}}{v_{ZZ}}|$ is the asymmetry parameter of the EFG tensor $V$. Since Sb2 site in CsCr$_3$Sb$_5$ has a similar environment as in CsV$_3$Sb$_5$, the asymmetry parameter $\eta$ should also be zero at high temperatures\cite{135-NMR}. By measuring the NMR frequency at different fields, the value of $\eta$ is indeed found to be zero from the slopes of $(f - f_0)/f_0$ vs. $f_0^{-2}$ (Fig.~\ref{fig:Ex_NMR}b in the Extended Data). All of these observations indicate that the NMR line being observed is originating from the Sb2 site.

The high-pressure experiments were carried out at the CAC station of SECUF. A standard four-probe method was used for resistivity measurements under high pressure in CAC. The sample was hung inside a Teflon capsule filled with glycerol pressure transmitting medium (PTM). The three-axis compression geometry together with the adoption of liquid PTM can ensure an excellent pressure homogeneity. The pressure values in CAC were estimated from the pressure-loading force calibration curve predetermined at room temperature. The mutual induction method was used for the ac magnetic susceptibility measurements in CAC. Several pieces of thin samples together with a piece of Pb served as the pressure marker and the superconducting reference were put inside the handmade primary/secondary coils of about 50 turns for each. The primary coil is driven by ac current of 1 mA and 317.7 Hz, while the output signal from the secondary coil was measured with a lock-in amplifier Stanford SR 830. Detail about the sample assembly and pressure calibrations of CAC can be found elsewhere\cite{CAC}. We also employed a piston-type high-pressure cell to finely tune the pressures in the range of 2 GPa $< P <$ 3 GPa.
Daphne 7373 was used as the pressure transmitting medium and, the pressure values were determined from the superconducting transition temperature of Pb according to the formula $P=(7.20-T_\mathrm{c}^{\mathrm{Pb}}$)/0.365 (GPa).

\begin{figure*}
\includegraphics[width=14cm]{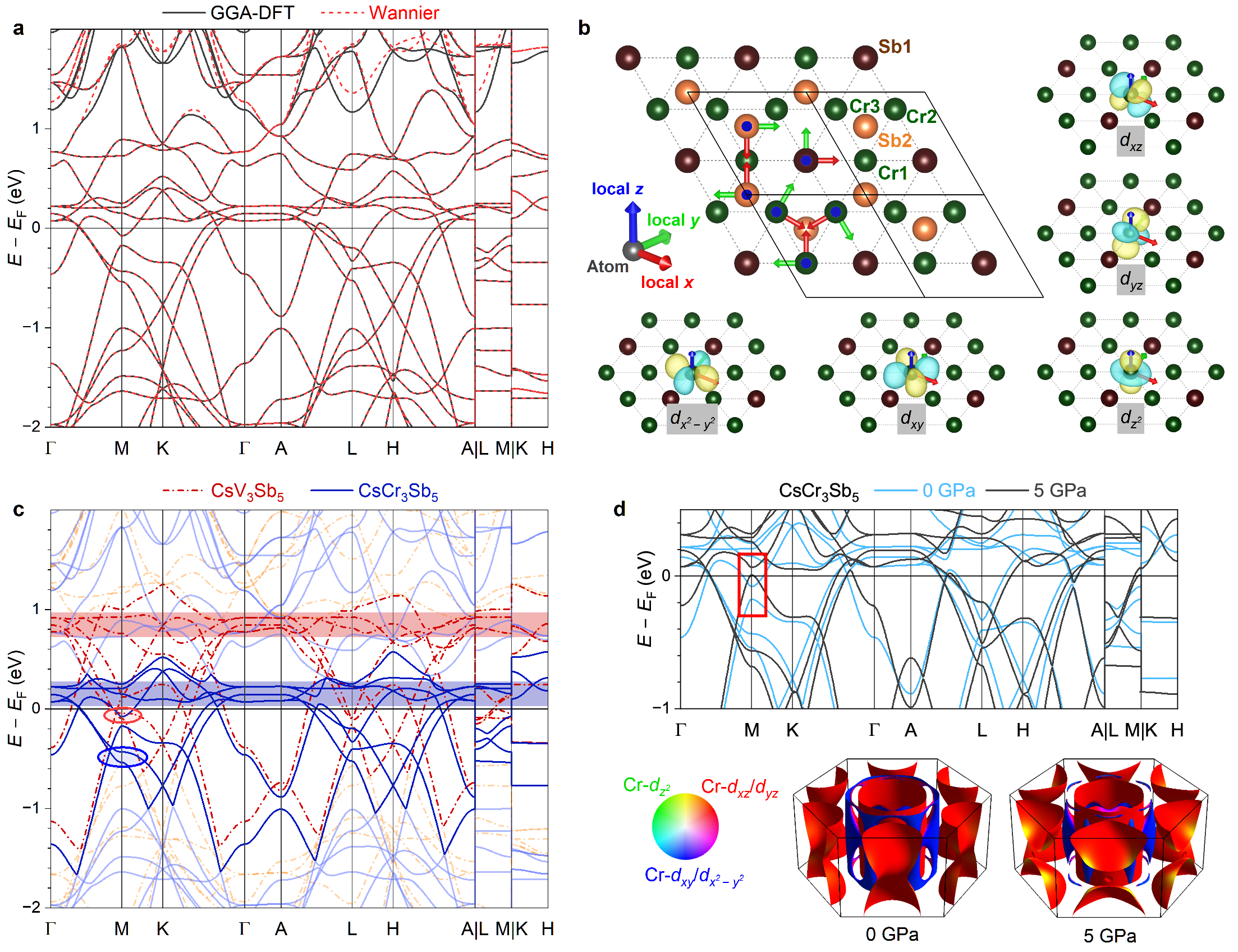}
\centering
\caption{\textbf{Extended Data Fig. 10 Additional information on the DFT calculations for nonmagnetic hexagonal CsCr$_3$Sb$_5$.} \textbf{a}, Band structures obtained from DFT calculations and from Wannier downfolding result. \textbf{b}, Local axis definitions and real-space distributions of Cr-3$d$ Wannier orbitals. \textbf{c}, Comparison of band structure with \CVS. The van Hove singularities and flat bands are highlighted with ellipses and transparent stripes, respectively. \textbf{d}, Comparison of band structure (top) and Fermi surface (bottom) at 0 and 5 GPa, respectively. Inside the red frame shows the main change in band structure. All the calculations were performed without spin-orbit interactions.
}
\label{fig:Ex_DFT}
\end{figure*}

\noindent\textbf{Electronic structure calculations}\\
The DFT-based first-principles calculations were performed using Vienna ab initio Simulation Package\cite{VASP}. The Kohn-Sham wave functions were treated with projected augmented wave method\cite{PAW}. The exchange-correlation energy was calculated with a Perdew-Burke-Ernzerhof (PBE) type functional\cite{GGA}. The energy cutoff of plane-wave basis was up to 500 eV and a $\Gamma$-centered 12$\times$12$\times$8 k-point mesh was employed in the self-consistent calculations. The experimental room-temperature crystal structure was adopted for the ambient-pressure calculations. As for the high-pressure calculations, the lattice constants and atomic coordinates were fully relaxed with a solid revised PBE (PBEsol) functional\cite{PBEsol}, leading to the calculated structural parameters of $a$ = 5.272 \AA, $c$ = 8.400 \AA, and $z(\mathrm{Sb2})=$ 0.2428.
The Fermi surface (plotted with FermiSurfer program\cite{FermiSurfer}) and orbital occupation results are calculated from our 30-band tight-binding Hamiltonians, which are obtained via Wannier downfolding\cite{Wannier}. As shown in Fig.~\ref{fig:Ex_DFT}a, the band structures around Fermi level are well reproduced with the Wannier downfolding results. During the Wannier downfolding processes, band disentanglements are performed with initial projectors of 15 Cr(V)-3$d$ and 15 Sb-5$p$ orbitals. The local coordinates are chosen so that the relative directions of Cr(V)-3$d$ orbitals are identical with respect to the 3-fold rotation axis (Fig.~\ref{fig:Ex_DFT}b).

\newpage
\section*{~}

\begin{table}
	\caption{Crystallographic data of CsCr$_3$Sb$_5$ at 298 K by the structural refinement of the single-crystal XRD. The unit of the equivalent isotropic and anisotropic displacement parameters is 0.001 $\mathrm{\AA}^2$.}\label{table1}%
	\begin{tabular*}{\textwidth}{@{\extracolsep\fill}ll}
		\hline
		Chemical formula &	CsCr$_3$Sb$_5$ \\
		Formula weight &	897.6 g/mol \\
		X-ray wavelength	& 0.71073 $\rm{{\AA}}$ \\
		Crystal system	& Hexagonal \\
		Space group	& $P$6/$mmm$ \\
		~ & $a$ = 5.4909(3) $\rm{{\AA}}$, $\alpha$ = 90$^\circ$ \\
		Unit-cell dimensions & $b$ = 5.4909(3) $\rm{{\AA}}$, $\beta$ = 90$^\circ$ \\
		~ & $c$ = 9.2417(5) $\rm{{\AA}}$, $\gamma$ = 120$^\circ$ \\
		Volume & 241.31(2) $\rm{\AA}^{3}$ \\
		$Z$ &	1 \\
		Density (calculated) & 6.1771 g/cm$^3$ \\
		Absorption coefficient & 20.646 mm$^{-1}$ \\
		$F$(000) & 382 \\
		Crystal size & 0.172  $\times$ 0.168  $\times$ 0.018 mm$^3$ \\
		$\theta$ range for data collection & 2.2$^\circ$ to 31.54$^\circ$ \\
		Index ranges & $-8 \leq$ \emph{h} $\leq$ 8, $-8 \leq$ \emph{k} $\leq$ 8, $-13 \leq$ \emph{l} $\leq$ 13 \\
		Reflections collected & 10522 \\
		Independent reflections & 207 [$R_{\mathrm{int}}$ = 0.0532] \\
		Completeness to $\theta$ = 31.54$^\circ$ & 100\% \\
		Refinement method & $F$ \\
		Data / restraints / parameters & 207 / 0 / 11 \\
		Goodness - of - fit	& 3.10 \\
		Final $R^{\ast}$ indices [$I$ $>$ 3$\sigma$($I$)]	& $R_{\rm{obs}}$ = 0.0283, $wR_{\rm{obs}}$ = 0.0467 \\
		$R$ indices [all data] & $R_{\rm{all}}$ = 0.0350, $wR_{\rm{all}}$ = 0.0477 \\
		Extinction coefficient & NA \\
		Largest diff. peak and hole	& 3.80 and $-$2.21 e$\rm{{\AA}}^{-3}$ \\
		\hline
	\end{tabular*}
\begin{tablenotes}\footnotesize
\item[1] $\ast$~ $R$ = $\Sigma\mid\mid$$F_o\mid -\mid$$F_c\mid\mid/\Sigma\mid F_o\mid$, $wR = {\Sigma[w(\mid F_o\mid^2 - \mid F_c\mid^2)^2] / \Sigma[w(\mid F_o\mid^4)]}^{1/2}$ and $w=1/(\sigma^2(F)+0.0001F^2)$.
\end{tablenotes}
~\\
\begin{tabular*}{\textwidth}{@{\extracolsep\fill}lccccccccccc}
				\hline
				Label &$x$&$y$&$z$ &Occupancy& $U_{\mathrm{eq}^*}$&	$U_{11}$&	$U_{22}$&	$U_{33}$&	$U_{12}$&	$U_{13}$&	$U_{23}$ \\
				Cs1 & 0&0&0&1&22(1)&21(1)&	21(1)&	23(1)&	10(1)&	0&	0\\
				Sb1	& 0&	0&	0.5000&	1&	15(1)&8(1)&	8(1)&	29(1)&	4(1)&	0&	0\\
				Sb2	&1/3&	2/3&	0.2624(1)&	1&	15(1)&15(1)&	15(1)&	16(1)&	7(1)&	0&	0 \\
				Cr1	&0.5&	0&	0.5&	1&	14(1)&12(1)&	15(2)&	17(2)&	7(1)&	0&	0\\
				\hline
\end{tabular*}
\begin{tablenotes}\footnotesize
\item[2] $\ast$~$U_\mathrm{eq}$ is defined as one third of the trace of the orthogonalized $U_{ij}$ tensor. The anisotropic displacement factor exponent takes the form: $-2\pi^2[h2a*2U_{11} + ...  + 2hka*b*U_{12}]$.
\end{tablenotes}
~\\
\begin{tabular*}{\textwidth}{@{\extracolsep\fill}ll}
\hline
Bond &	Distance ($\rm{\AA}$)~~~~~~~~~~~~~~~~~~~~~~~~~~~~~~~~~~~~~~~~~~~~~~\\
Cr-Cr	&2.7455(5)$\times$4\\
Cr-Sb	&2.7455(3)$\times$2 \\
Cr-Sb & 2.7081(7)$\times$4 \\
Cs-Sb	&3.9914(6)$\times$12 \\
\hline
\end{tabular*}
\end{table}

\newpage
\begin{table}
\renewcommand{\arraystretch}{1.6}
\footnotesize
\centering
\caption{Comparison of structural and physical properties of Cs$M$$_3$Sb$_5$~($M$ = V and Cr). For simplicity, V135 and Cr135 denote \CVS~ and \CCS, respectively, where the space is limited.}
\label{comparison}  
\tabcolsep=1pt
\scalebox{1}{
\begin{tabular*}{\textwidth}{@{\extracolsep\fill}lllll}
\toprule[0.25mm]
~& Parameters & \multirow{1}{6em}{CsV$_3$Sb$_5$} & \multirow{1}{3em}{CsCr$_3$Sb$_5$} & \multirow{1}{18em}{Notes}\\
\hline
\multirow{3}{8em}{Crystal structure at room temperature} & $a$~($\rm{{\AA}}$) &  5.5236(6)~[Ref.{\cite{2023PRM.structure}}] & 5.4909(3) & $\Delta a/a_\mathrm{V} = -$0.59\%\\
& $c$~($\rm{{\AA}}$) &  9.3623(15)~[Ref.{\cite{2023PRM.structure}}] & 9.2417(5) & $\Delta c/c_\mathrm{V} = -$1.3\% \\
& $d_{M-M}$ ($\rm{{\AA}}$) & 2.762(4)~[Ref.{\cite{2023PRM.structure}}] &  2.7455(5) & $\Delta d = -$0.0164~($\rm{{\AA}}$) \\
\hline
\multirow{4}{8em}{Phase transition at ambient pressure} & $T_{\mathrm{DW}}$~(K)  &94~[Ref.{\cite{2020PRL.Cs135SC}}] & 55 & \multirow{4}{10em}{CDW transition for \CVS;\\
AFM CDW/SDW-like transition for \CCS}\\
&In-plane modulation& $2\times2$~[Ref.\cite{2022NSR.review}]&$4\times1$ &\multirow{3}{4em}{}\\
&Out-of-plane modulation& $\times2$ or $\times4$~[Ref.\cite{2022NSR.review}]& None&\\
&Structural distortion& orthorhombic~[Ref.\cite{2023PRM.structure}]& Monoclinic&\\

\hline
\multirow{10}{8em}{Physical properties at ambient pressure} & $\rho_{ab}^{300\rm{K}}$~(m$\Omega$~cm) & 0.3~[Ref.{\cite{rou_c_V135}}] & 1.43 & Bad metal for \CCS\\
& $\rho_{c}^{300\rm{K}}$~(m$\Omega$~cm) & 3.2~[Ref.{\cite{rou_c_V135}}] & 81.5 & ($\rho_{c}/\rho_{ab}$)$^{3\rm{K}}_{\rm{V135}}$ = 20 \\
& $\rho_{ab}^{3\rm{K}}$~(m$\Omega$~cm) & 0.005~[Ref.{\cite{rou_c_V135}}] & 0.85 & ($\rho_{c}/\rho_{ab}$)$^{3\rm{K}}_{\rm{Cr135}}$ = 85 \\
& $\rho_{c}^{3\rm{K}}$~(m$\Omega$~cm) & 0.1~[Ref.{\cite{rou_c_V135}}] & 72.3 & ~\\
& $\chi^{300\rm{K}}$~(10$^{-3} $emu/mol) & 0.37~[Ref.{\cite{2020PRL.Cs135SC}}] & $\sim$2 & $\chi_{\rm{Cr}}$/$\chi_{\rm{V}}$ = 5.4\\
&\multirow{1}{10em}{Effective moment ($\mu_{\mathrm{B}}$/Cr)}&\multirow{1}{8em}{None [Ref.{\cite{2020PRL.Cs135SC}}]}&1.26$\pm$0.12 & CW paramagnetic for Cr135\\

& \multirow{2}{12em}{$D(E_{\mathrm{F}})$ (states eV$^{-1}$~fu$^{-1}$)} & \multirow{2}{8em}{10~[Ref.{\cite{2020PRL.Cs135SC}}]} & \multirow{2}{8em}{7.7~(DFT)} & \multirow{1}{18em}{$E_\mathrm{F}\rightarrow$van Hove for V135} \\
& &&&\multirow{1}{18em}{$E_\mathrm{F}\rightarrow$ Flatbands for Cr135} \\
& $\gamma$~(mJ K$^{-2}$ mol$^{-1}$) & 20.03~[Ref.{\cite{2021SCP.YHQ}}] & 105 & Large electron mass for Cr135\\
\hline
\multirow{4}{8em}{Superconductivity}& $T_{\mathrm{c}}^{\rm{AP}}$~(K)$^\ast$  &2.5~[Ref.{\cite{2020PRL.Cs135SC}}] & None & Absence of AP SC for Cr135\\
& $T_{\mathrm{c},\rm{max}}^{\rm{HP}}$~(K)$^\ast$  &8@2~GPa~[Ref.{\cite{2021PRL.HP.ChengJG}}] & 6.4@4.2~GPa & Possible USC for Cr135\\
& $H_{\rm{c2}}^{\rm{max}}$(0)~(T)  &3.5~[Ref.{\cite{2021PRL.HP.ChengJG}}] & 14.34 & \multirow{2}{14em}{Exceeding the Pauli-paramagnetic limit $H_\mathrm{P}$ for Cr135}\\
& $\mid$d$H_{\rm{c2}}^{\rm{max}}$/d$T$($T_{\mathrm{c}}$)$\mid$~(T/K) & 0.58~[Ref.{\cite{2021PRL.HP.ChengJG}}] & 5.03 & \\
\bottomrule[0.25mm]
\end{tabular*}}
\begin{tablenotes}\footnotesize
\item[4] $\ast$ AP and HP denote ambient pressure and high pressure, respectively.
\end{tablenotes}
\end{table}

\section*{~}

\end{document}